\documentclass[a4paper,11pt]{article}
\usepackage{lineno}
\usepackage[subrefformat=parens,labelformat=parens]{subfig}
\usepackage{xspace}
\usepackage{pos}

\title{Performance of a proposed event-type based analysis for the Cherenkov Telescope Array}
 \ShortTitle{Performance of a proposed event-type based analysis for the CTA}

\author*[a]{T. Hassan}
\author[b]{O. Gueta}
\author[b]{G. Maier}
\author[c]{M. Nöthe}
\author[d]{M. Peresano}
\author[e]{I. Vovk}

\affiliation[a]{Centro de Investigaciones Energéticas, Medioambientales y Tecnológicas (CIEMAT), Av. Complutense, 40, 28040 Madrid, Spain}

\affiliation[b]{DESY, Platanenallee 6, 15738 Zeuthen, Germany}
\affiliation[c]{Department of Physics, TU Dortmund University, Otto-Hahn-Str. 4a, 44227 Dortmund, Germany}
\affiliation[d]{AIM, CEA, CNRS, Universite Paris-Saclay, Universite Paris Diderot, Sorbonne Paris Cite, F-91191 Gif-sur-Yvette, France}
\affiliation[e]{Institute for Cosmic Ray Research, The University of Tokyo
5-1-5 Kashiwa-no-Ha, Kashiwa City, Chiba, 277-8582, Japan}

\forColl{CTA Consortium} 

\emailAdd{tarek.hassan@ciemat.es}
\emailAdd{orel.gueta@desy.de}

\abstract{The Cherenkov Telescope Array (CTA) will be the next-generation observatory in the field of very-high-energy (20 GeV to 300 TeV) gamma-ray astroparticle physics. Classically, data analysis in the field maximizes sensitivity by applying quality cuts on the data acquired. These cuts, optimized using Monte Carlo simulations, select higher quality events from the initial dataset. Subsequent steps of the analysis typically use the surviving events to calculate one set of instrument response functions (IRFs). An alternative approach is the use of event types, as implemented in experiments such as the Fermi-LAT.  In this approach, events are divided into sub-samples based on their reconstruction quality, and a set of IRFs is calculated for each sub-sample. The sub-samples are then combined in a joint analysis, treating them as independent observations. This leads to an improvement in performance parameters such as sensitivity, angular and energy resolution. Data loss is reduced since lower quality events are included in the analysis as well, rather than discarded. In this study, machine learning methods will be used to classify events according to their expected angular reconstruction quality. We will report the impact on CTA high-level performance when applying such an event-type classification, compared to the classical procedure.}

\FullConference{37$^{\rm{th}}$ International Cosmic Ray Conference (ICRC 2021)\\
		July 12th -- 23rd, 2021\\
		Online -- Berlin, Germany}

\begin{document}
\maketitle

\section{Introduction}

The Cherenkov Telescope Array\footnote{\url{www.cta-observatory.org}} (CTA) will be a next-generation observatory employing an array of imaging atmospheric Cherenkov telescopes (IACTs). The observatory will be built in two different sites, one in each hemisphere. It will provide a major improvement with respect to the current generation of IACTs both in sensitivity and in angular and energy resolution over a very broad energy range (20 GeV up to more than 300 TeV). This improvement will be possible with a cost-effective solution employing arrays of IACTs coming in three different sizes: large-sized telescopes (LSTs, 23-m diameter), medium-sized telescopes (MSTs, 11.5-m diameter) and small-sized telescopes (SSTs, 4.3-m diameter).

The future performance of CTA is estimated from detailed Monte Carlo (MC) simulations. It is encapsulated in a set of Instrument Response Functions (IRFs) such as effective area, energy or angular resolution or residual background rate. The calculation, comparison and ranking of these IRFs and other associated figures of merit have been key in assessing the scientific prospects of CTA, in guiding the development of the different telescope designs\footnote{\url{https://www.cta-observatory.org/project/technology/}}, in choosing the CTA sites and in fixing the array layouts~\cite{CTAScience, CTAsites, CTAarrays}. The methodology used to derive the performance of the future CTA, i.e., the computation of its expected sensitivity and associated IRFs, has been widely described in previous contributions (see e.g., Refs.~\cite{CTAsites, CTAarrays}) and is briefly discussed in section~\ref{sec:optimization}. 

As proven by the success of the periodic data releases performed by the \textit{Fermi} Large Area Telescope (LAT) Collaboration~\cite{fermiLAT}, high-level analysis performance can be significantly boosted by improving the event selection, reflecting the knowledge we have on the performance of the detector into the derived IRFs. By partitioning \textit{Fermi}-LAT events into different Point Spread Functions (PSF) event types, and by computing sets of IRFs specific to each of these types, high-level analysis tools are able to include the extra knowledge provided within the IRFs of the different quality of each event into the likelihood analysis~\cite{fermipy, latpass8}. The benefits achieved by this event-type partitioning range from reducing background contamination to increasing the effective area (and therefore sensitivity) and improving the angular and energy resolution for the subset of the highest quality events.

In this contribution we study the performance of a similar event-type partitioning scheme for CTA data analysis. We test an event partitioning equivalent to the PSF event type used by  \textit{Fermi}-LAT, and explore the benefits and drawbacks of such an approach.

\section{Cut optimization and performance evaluation}
\label{sec:optimization}

Detailed MC simulations are generated and used to derive the expected response of CTA telescopes to very-high-energy gammas, as well as protons and electrons (the particles which constitute the main irreducible background limiting CTA performance)~\cite{corsika, Simtelarray}. A classical IACT analysis is performed on these MC samples, similar to the data analysis employed by the current generation of IACTs~\cite{Eventdisplay, MARS}. The product of this analysis, generally referred to as Data Level 2 (DL2), contains the reconstructed direction, energy and the likelihood to be signal (gamma-like) or background (mainly proton-like) of each analyzed event. These DL2 tables are re-weighted to resemble the particle statistics expected from standard CTA observations on a Crab-Nebula-like source, and a cut optimization procedure is performed to find the cuts maximizing the sensitivity of the array as a function of the reconstructed energy. This process specifically optimizes the size of the selected ON region (the region defined as the signal region), event multiplicity (number of CTA telescopes simultaneously detecting and reconstructing each event) and the background rejection likelihood. The optimised cuts are used to produce a single set of IRFs. 

This procedure has certain limitations: 1) The amount of data actually used (those events surviving these cuts) is relatively small compared with the original dataset. This leads to a considerable amount of data being rejected, that could nevertheless be useful. 2) Once IRFs are computed, all the extra knowledge we have in lower-level analysis steps (for instance, on the expected quality of each individual event) is lost and, from that point on, all events surviving quality cuts are treated equally. 

Event-type partitioning, as proven by the \textit{Fermi}-LAT Collaboration~\cite{latpass8}, has the potential of better reflecting the knowledge we have on the reconstruction of each event, and of using that information to improve the high-level performance of CTA analysis. In the case of CTA, there are certain parameters that are known to reflect the reconstruction quality of the events. For instance, event multiplicity provides information that is known to dramatically affect the angular resolution of the events, but is lost if a single set of IRFs is produced.

\section{Event-type partitioning and data analysis}
\label{sec:analysis}

The methodology described in the previous section was modified in the following manner to implement a PSF event-type partitioning:
\begin{itemize}
    \item The starting data products are the result of the low-level analysis, i.e., DL2 tables.
    \item Gamma-ray DL2 tables are divided into 3 different samples: training sample, test sample, and IRF-production sample.
    \item A regression machine learning algorithm is trained on the training sample to predict for each event the angular reconstruction quality $log \Delta d$ (the difference between the simulated and reconstructed direction of each event, in logarithmic scale). The performance of the machine learning algorithm is evaluated with the test sample.
    \item The trained regressor is applied to the gamma-ray IRF-production sample and to the rest of the DL2 tables involved in the performance evaluation (protons and electrons), predicting the angular reconstruction quality of each event.
    \item Gamma-ray DL2 events (from the IRF-production sample) are ranked according to their predicted angular reconstruction quality in equally spaced steps in logarithmic reconstructed energy, and separated into N event type samples with an equal number of events. The angular reconstruction quality thresholds\footnote{Specific threshold values of $log \Delta d_i$ as a function of the energy, used to define the edges of each of the N event types in the partitioning stage.} are defined with this sample, and applied to proton and electron DL2 tables, also separated into N samples.
    \item An identical cut optimization and IRF computation is performed on each of the N independent DL2 tables (gamma, proton and electron), following an identical methodology as described in section~\ref{sec:optimization}.
\end{itemize}

All of the code used for this project is available under an open-source BSD-3 license~\cite{event_types_github}. For the results presented here, we used DL2 analysis results from the EventDisplay analysis chain~\cite{Eventdisplay} from the fifth large-scale CTA MC production~\cite{prod-5}. The DL2 tables contain simulated observations at 20 degrees in zenith pointing both to North and South directions. Diffuse gamma-ray simulations\footnote{Arrival directions of simulated gamma rays cover the whole field of view of CTA telescopes} were used for the training and test samples (with a 75\% to 25\% ratio, respectively), while point-like gamma-ray simulations\footnote{Simulated gamma rays come from a single point at the centre of the field of view of CTA telescopes} were used for the IRF-production sample (full available statistics). A wide variety of machine learning algorithms were tested, both from the \textit{sklearn} and \textit{tensorflow} libraries \cite{sklearn, tensorflow}. A long list of low-level training features (43 in total) were used during the training \cite{event_types_github}. We perform the cut optimization and produce a set of IRFs with \textit{pyirf}~\cite{pyirf}. 

\section{Results}

By following the methodology discussed in section~\ref{sec:analysis}, we report in the following subsections on the results of both the optimization effort to maximize the angular performance prediction and the resulting CTA performance estimations. 

\subsection{Performance of angular reconstruction quality predictors}

\begin{figure}
\centering
\includegraphics[width=0.48\linewidth]{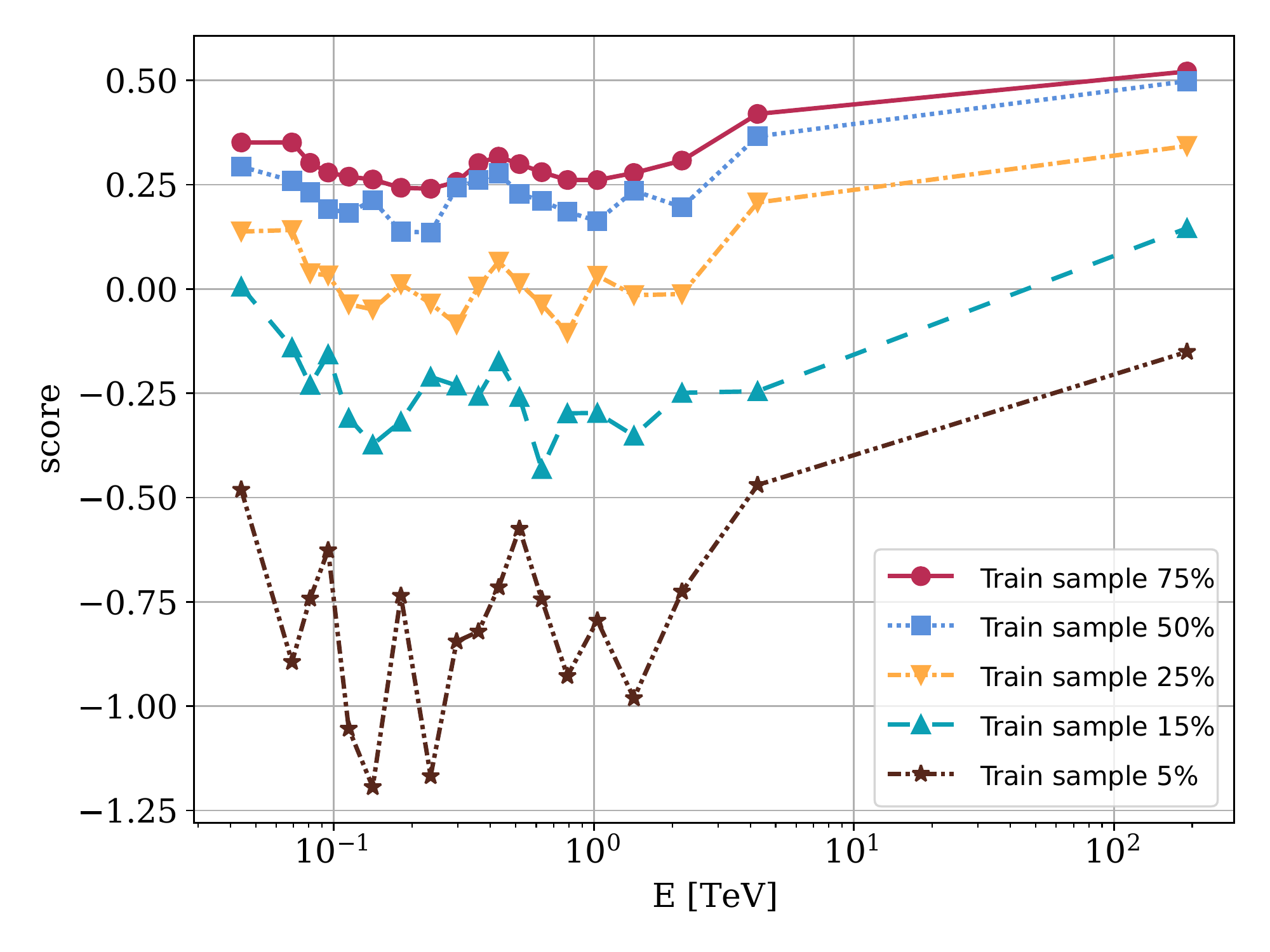}
\\
\includegraphics[width=0.4\linewidth]{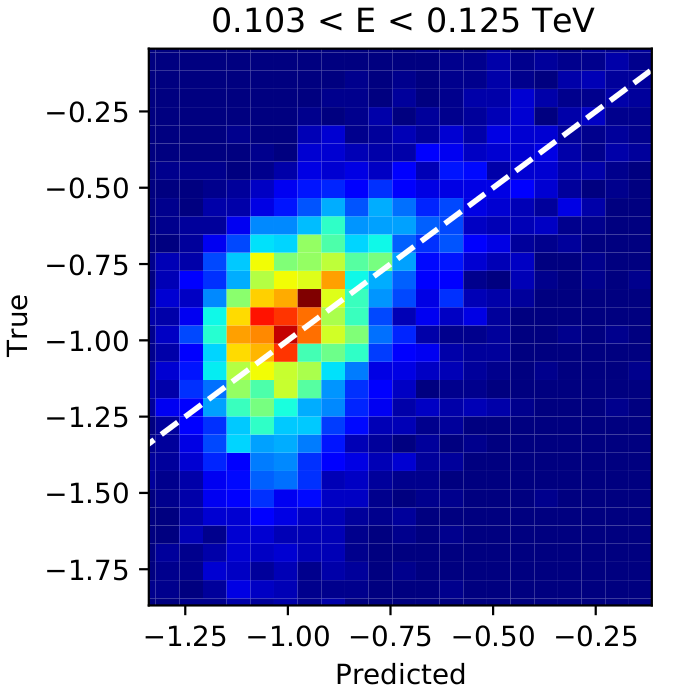}
\includegraphics[width=0.4\linewidth]{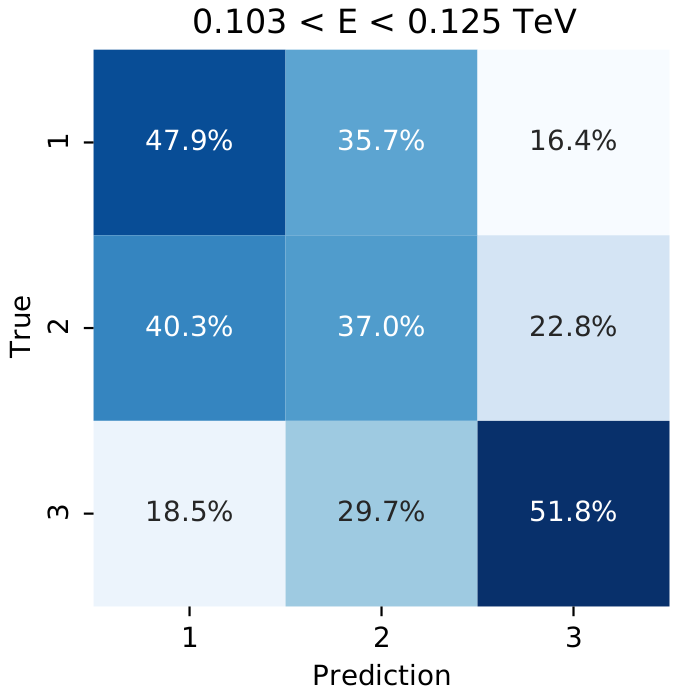}
\caption{\label{fig:ML_performance}
\textit{Top}) Performance score vs reconstructed energy for different training statistics. Label refers to the percentage of the training sample with respect to the total available training sample. A higher score indicates better performance. \textit{Bottom left}) As an example, true vs predicted distribution of the angular reconstruction quality of a single energy bin, defined as the logarithmic difference between simulated to reconstructed event direction. \textit{Bottom right}) confusion matrix for the same energy range, obtained using a 3 event-type partitioning with equal statistics.}
\end{figure}

Predictions were performed using several regression machine learning algorithms. The reason why classification methods were not used was that regressors provide similar performance while allowing a better control of the subsequent event-type partitioning. A preliminary evaluation suggests that the best performance is provided by a multilayer perceptron (MLP) neural network with a $\textit{tanh}$ as neuron activation function. 

The required training statistics to reach good performance is a key aspect of this study, as CTA computing requirements could be the limiting factor to realistically implement the methods proposed in this work. Fig.~\ref{fig:ML_performance} (top) shows that available diffuse gamma-ray simulations statistics is probably not enough to reach the best performance possible, as ideally the same simulations would be needed to compute IRFs over the whole field of view of CTA. For the results shown here, this is not an issue, as we use an independent point-like gamma-ray sample to compute PSF event-types performance.

As an example, for a data separation into $N=3$ PSF event types, fig.~\ref{fig:ML_performance} (bottom left and right) shows the confusion matrix\footnote{Matrix showing the percentage of correct/incorrect classifications of each event type by the algorithm.} on both the actual distribution of the angular reconstruction quality and event-type classification of a single energy bin. For this specific energy range, we can see how "good" events (type 1) are mostly predicted into the first two event types, while "bad" events are generally well characterized within event type 2 or 3.

\subsection{CTA PSF event type performance}

Following the methodology described in the previous sections, we apply a PSF event partitioning into 3 different event types (mainly limited by the background MC statistics) and calculate standard CTA IRFs for a point-source located at the centre of the field of view in 50 hours of observing time for one of the potential layouts of the southern array (14 MSTs and 40 SSTs), as in~\cite{prod-5}. 

\begin{figure}
\centering
\includegraphics[width=0.48\linewidth]{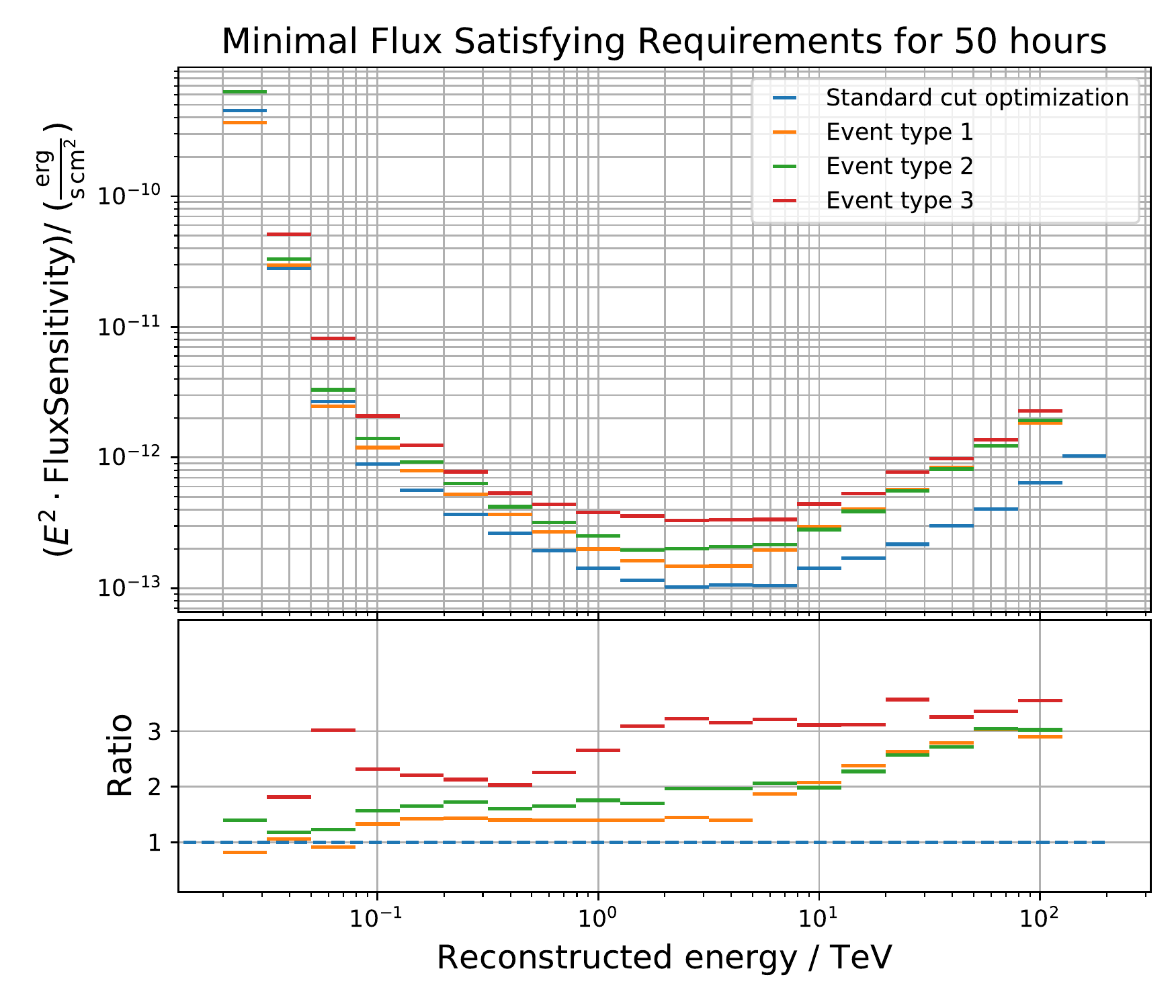}
\includegraphics[width=0.48\linewidth]{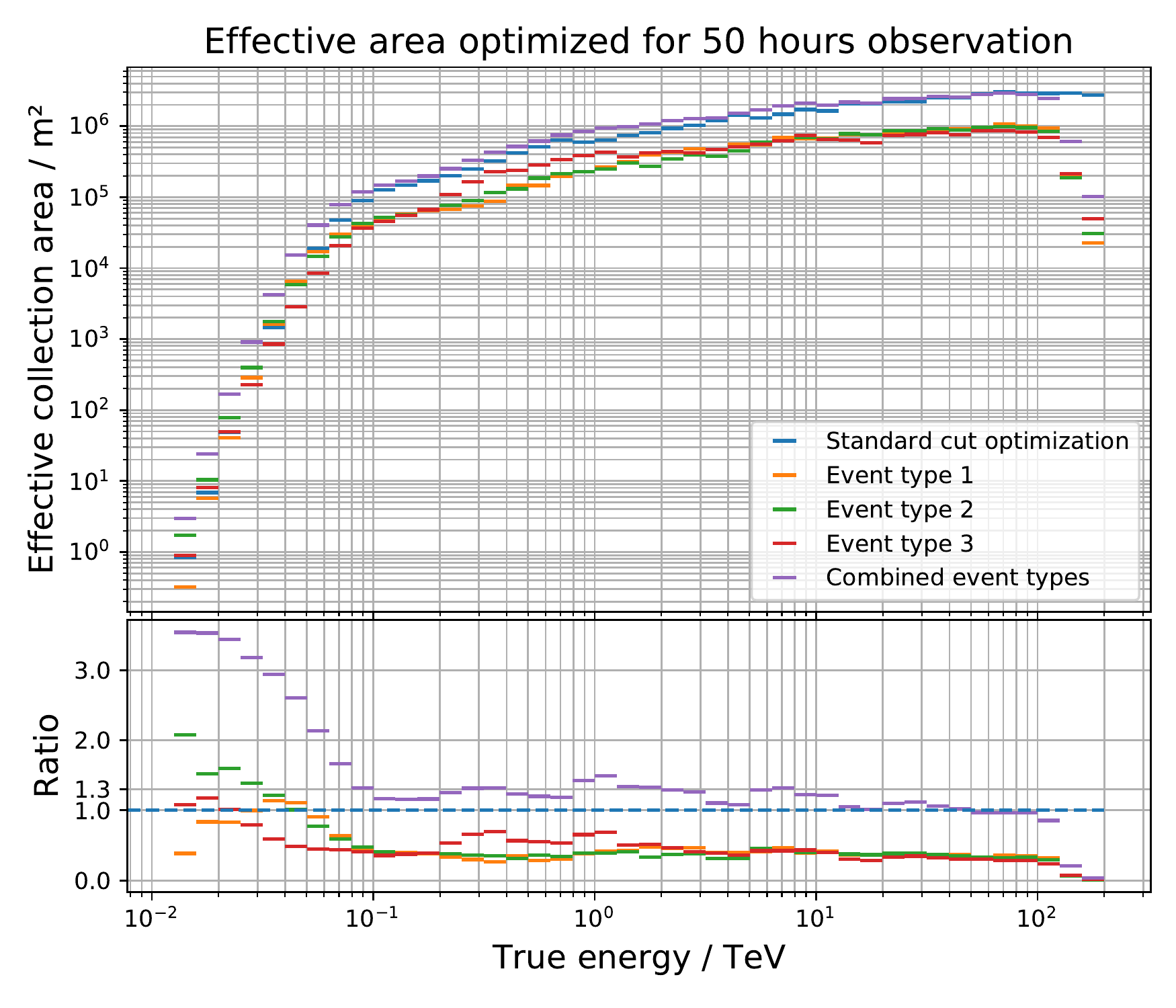}
\includegraphics[width=0.48\linewidth]{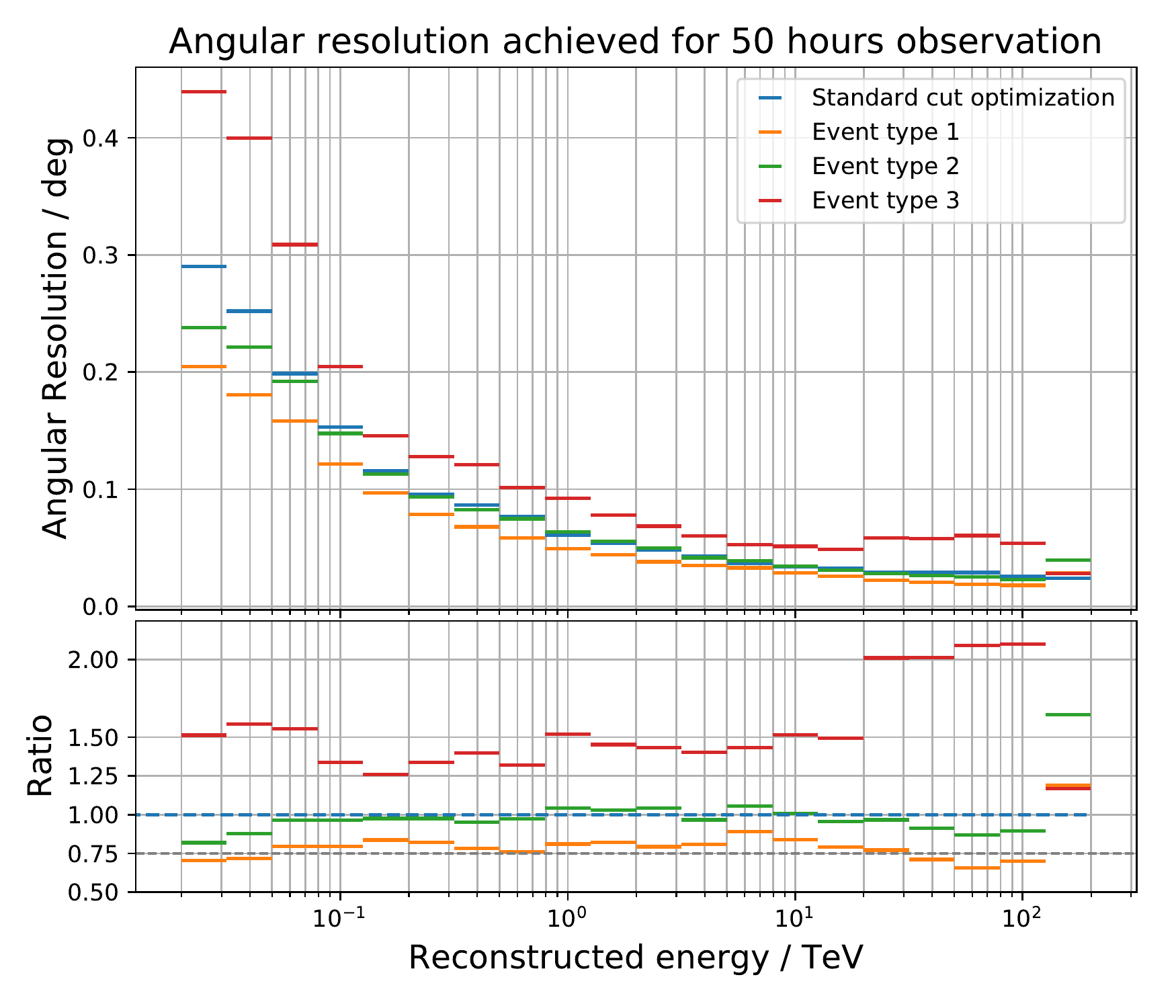}
\includegraphics[width=0.48\linewidth]{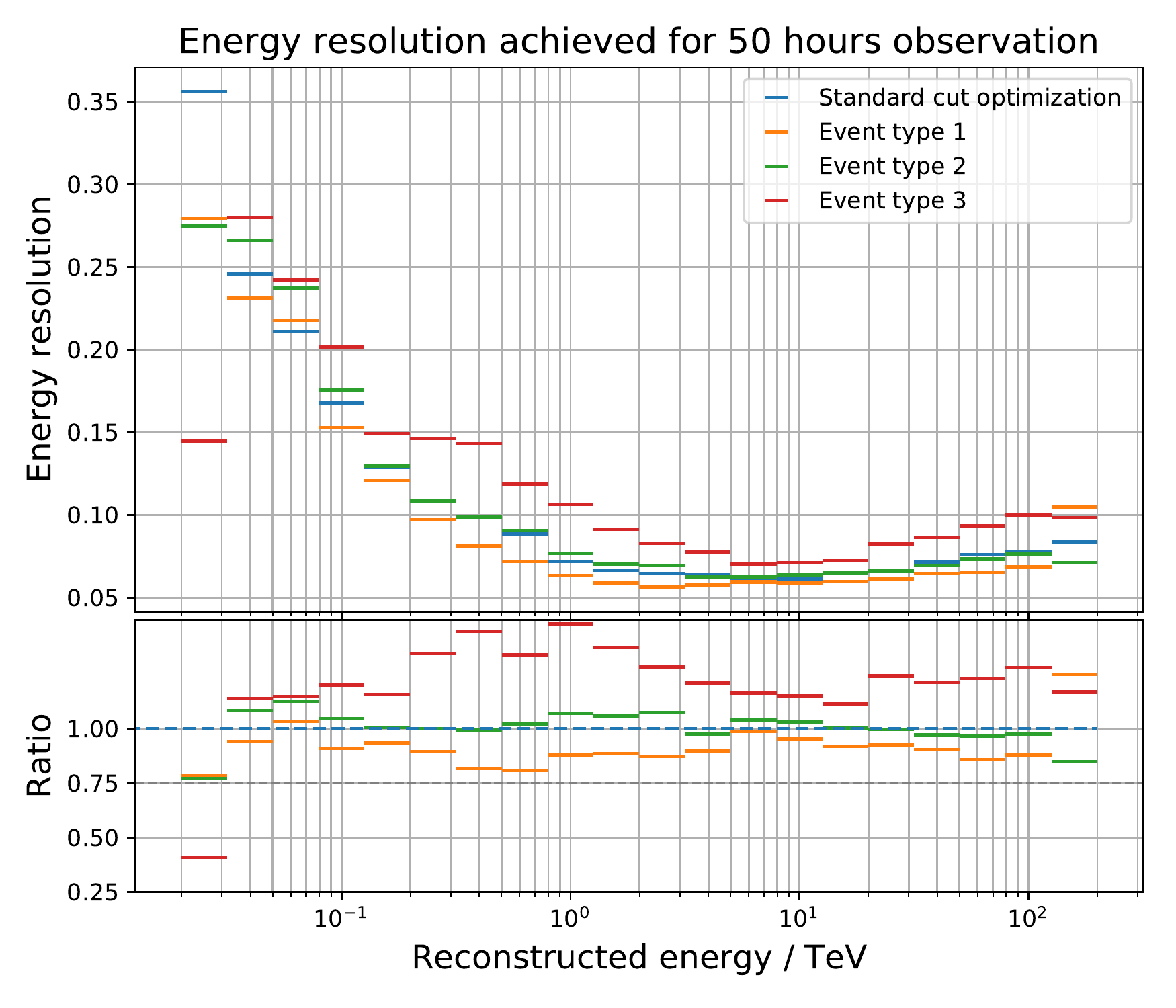}
\caption{\label{fig:CTAperformance}
\textit{Top left}) Sensitivity vs reconstructed energy of a potential layout for the southern array (14 MSTs and 40 SSTs) for a point-source located at the centre of the field of view for 50 hours of observation time. Note worse sensitivity of each event type sub-sample is expected, as 33$\%$ of the events are used in each of them. \textit{Top right}) Effective area vs true energy for the same array and conditions. The drop appearing at the highest energies is just an effect of lacking proton MC statistics when dividing the sample into 3 event types. \textit{Bottom left}) Angular resolution vs reconstructed energy for the same array and conditions. \textit{Bottom right}) Energy resolution vs reconstructed energy for the same array and conditions.}
\end{figure}

As shown in fig.~\ref{fig:CTAperformance} (top left), the comparison between the standard single-IRF cut optimization and the event-type-wise IRFs is not trivial in terms of sensitivity. This is expected given that each event-type-wise IRF only contains 33$\%$ of the sample. When looking to the effective area in fig.~\ref{fig:CTAperformance} (top right), we see the amount of event statistics actually used when combining the 3 defined event types is roughly 3.5 times larger for the lowest energies, while in the core energy range of CTA the extra statistics is roughly 25$\%$.

Given the event-type partitioning focused on the PSF, angular resolution is the most relevant figure of merit of this study. Fig. \ref{fig:CTAperformance} (bottom left) shows the PSF event-type partitioning is indeed providing an improved angular resolution of roughly 25$\%$ over all energies for the top 33$\%$ of the classified events. The event type 2 provides roughly equivalent angular resolution as the one resulting from the standard cut optimization, while the event type 3 clearly identifies those events that have a worse reconstruction across all energies.

Even if it was not the focus of this work, angular and energy resolution are known to be highly correlated within IACTs low-level analysis, i.e. those events with better angular reconstruction are generally also expected to have better energy reconstruction. For this reason, it also makes sense to check how the resulting energy resolution looks, even if the optimized parameter was the angular reconstruction. As shown in fig.~\ref{fig:CTAperformance} (bottom right), event type 1 does indeed also provide an improved energy resolution, event type 2 has the energy resolution roughly resembling the one calculated within standard IRFs, while event type 3 properly identifies those events with clearly worse energy resolution.

\section{Conclusions}

Applying an event-type partitioning prior to the IRF computation has been extremely successful for the \textit{Fermi}-LAT analysis. In this work we demonstrate that it also shows great potential for improving CTA future capabilities. By training machine learning methods to predict the angular reconstruction quality of CTA simulated events, we show we are able to separate events according to their quality, providing a 25$\%$ improved PSF across all energies for a sub-sample of the events. This classification also provides improved energy resolution, although following an identical methodology one could repeat the study focusing on improving energy resolution (as done in fact by \textit{Fermi}-LAT). This improvement both in angular and energy resolution has also been achieved when applying the described event-type analysis to the one of the candidate CTA northern array layouts..

The resulting event-type-wise sensitivities hint towards a potential improvement at the lowest energies of CTA, coming from the improved cut optimization, as we are indeed providing more information to the cut optimization process by dividing the dataset into samples of different quality. A full high-level simulation of a Crab-Nebula-like observation is needed to test if PSF event types indeed provide a net gain in sensitivity. 

The fact that the angular and energy reconstruction are highly correlated within IACTs analysis may actually present a problem for high-level data analysis. Accounting for such a correlation could be computationally prohibitive, while not taking the correlation into account would lead to unrealistic results. By separating events into different event types, the effect of this correlation on the high-level analysis is highly suppressed, as the different IRFs from different event types would represent more realistically the performance of those events.

Applying event-type partitioning comes at a cost: additional MC statistics could be needed to provide an independent sample for the training/testing of the methods, and also because the computation of an increased number of IRFs may require additional data to reach similar statistical uncertainties. This investigation shows that even if additional MC statistics may indeed be required, standard angular/energy reconstruction methods \cite{ctapipe} may already be capable of predicting event-wise expected performance (therefore not requiring an independent data sample for their training). Regarding the event statistics required for the IRF computation, we have seen it will not produce an enormous extra stress on CTA computing requirements as IRFs will be computed mainly from events that would be in any case discarded.

\section*{Acknowledgments}

\noindent This work was conducted in the context of the CTA Consortium and CTA Observatory.

\noindent We gratefully acknowledge financial support from the agencies and organizations listed here: 
\url{http://www.cta-observatory.org/consortium_acknowledgments}.

\clearpage
\section*{Full Authors List: \Coll\ }


\scriptsize
\noindent
H.~Abdalla\textsuperscript{1}, H.~Abe\textsuperscript{2},
S.~Abe\textsuperscript{2}, A.~Abusleme\textsuperscript{3},
F.~Acero\textsuperscript{4}, A.~Acharyya\textsuperscript{5}, V.~Acín
Portella\textsuperscript{6}, K.~Ackley\textsuperscript{7},
R.~Adam\textsuperscript{8}, C.~Adams\textsuperscript{9}
% ...
\\

\noindent

1 : Centre for Space Research, North-West University, Potchefstroom, 2520, South Africa

2 : Institute for Cosmic Ray Research, University of Tokyo, 5-1-5, Kashiwa-no-ha, Kashiwa, Chiba 277-8582, Japan

3 : Pontificia Universidad Católica de Chile, Av. Libertador Bernardo O'Higgins 340, Santiago, Chile

4 : AIM, CEA, CNRS, Université Paris-Saclay, Université Paris Diderot, Sorbonne Paris Cité, CEA Paris-Saclay, IRFU/DAp, Bat 709, Orme des Merisiers, 91191 Gif-sur-Yvette, France

5 : Centre for Advanced Instrumentation, Dept. of Physics, Durham University, South Road, Durham DH1 3LE, United Kingdom

6 : Port d'Informació Científica, Edifici D, Carrer de l'Albareda, 08193 Bellaterrra (Cerdanyola del Vallès), Spain

7 : School of Physics and Astronomy, Monash University, Melbourne, Victoria 3800, Australia

8 : Laboratoire Leprince-Ringuet, École Polytechnique (UMR 7638, CNRS/IN2P3, Institut Polytechnique de Paris), 91128 Palaiseau, France

9 : Department of Physics, Columbia University, 538 West 120th Street, New York, NY 10027, USA

% ...

\end{document}